\documentclass[prd,onecolumn,preprint,nofootinbib]{revtex4}
\usepackage[plainpages=false, colorlinks=true, anchorcolor=blue, linkcolor=blue, citecolor=blue, bookmarks=false]{hyperref}
\usepackage{graphicx}% Include figure files
\usepackage{dcolumn}% Align table columns on decimal point
\usepackage{bm}% bold math
\usepackage{amsmath}
\usepackage{appendix}
\usepackage{subcaption}
\newcommand{\rthis}[1]{\textcolor{black}{#1}}
\begin{document}

%\preprint{APS/123-QED}

\title{\textbf{Limit on high energy neutrino emission from Abell 119 using IceCube 10-year muon track data}}% 

\author{Sri Devaki \surname{Meduri}}
 \altaffiliation{E-mail: ep23btech11030@iith.ac.in}
\author{Shantanu Desai}%
 \altaffiliation{E-mail: shntn05@gmail.com}

\affiliation{Department  of Physics, IIT Hyderabad,  Kandi, Telangana-502284, India}

\begin{abstract}
We carry out a search for high energy muon neutrino emission from the galaxy cluster Abell 119, motivated by a recent \rthis{tentative} detection of GeV 
gamma-ray emission  from this cluster using the Fermi-LAT telescope, which hinted at a hadronic origin. For this purpose, we used the 10-year muon track data from 2008-2018, provided by the IceCube Collaboration and implement the unbinned maximum likelihood method. We do not find any statistically significant excess and the observed value of the test statistic is consistent with a null result. We then obtain upper limits (at 90\% confidence level) on the differential muon  neutrino energy flux from this cluster, whose value is equal to $1.2 \times 10^{-9}~\mathrm{GeV}~\mathrm{cm}^{-2}~\mathrm{s}^{-1}$ at a pivot energy of  100 TeV, after assuming a neutrino spectral index of \rthis{2.0}. The neutrino flux expected under the hadronic interpretation is,  however, approximately  an order of magnitude smaller than our flux limit, and hence the hadronic model cannot be ruled out based on our upper limit. Among the next generation neutrino detectors, \rthis{IceCube-Gen2} and TRIDENT neutrino detector should be able to confirm or rule out a hadronic origin for the gamma-ray emission in Abell 119 with 10 years of exposure.

\end{abstract}

\maketitle

%\tableofcontents

\section{Introduction}
\label{sec:intro}
In 2013, IceCube detected a diffuse neutrino flux observed in the TeV-PeV energy
range~\cite{IceCubescience}, whose origin is still unknown~\cite{Halzen23}.  The first point source for which a statistically significant spatial \rthis{and temporal} association  was found was the blazar TXS 0506+056~\cite{IceCubeblazar}.
\rthis{The statistical significance of this spatial and temporal association with this blazar using 10 years of IceCube muon track data has been estimated to be $3.5\sigma$ and $2.4\sigma$, respectively~\cite{IceCubedata}}.
The second such source with a  significance of $4.2\sigma$ was NGC 1068, a Seyfert galaxy~\cite{IceCubeSeyfert}. However, these sources can only account for up to  1\% of the diffuse astrophysical neutrino flux~\cite{IceCubedata,IceCubeSeyfert,IceCube20}. 
In addition to the diffuse neutrino flux,  the IceCube Collaboration  also recently obtained   4.5$\sigma$ evidence for neutrino emission from the Galactic plane using  cascade events~\cite{Science}. This signal is consistent with diffuse emission from the galactic plane, although a contribution from  a population of unresolved point sources cannot be ruled out~\cite{Science}.

Therefore, considerable effort has been devoted both within and outside the IceCube collaboration to determine the origin of the diffuse neutrino flux. Recently the IceCube Collaboration identified additional X-ray bright Seyfert galaxies  such as NGC 4151, CGCG 420-015 (in addition to NGC 1068)  as promising sources~\cite{IceCubeXrayseyfert,IceCubeSeyfert2}. Most recently, a 3.3$\sigma$ excess was found from 11 such Seyfert galaxies (which excludes NGC 1068)~\cite{IceCubeSeyfert2}. Outside the IceCube Collaboration, evidence for $3\sigma$ excess has recently been found from a stacked search involving 683 supermassive black hole binaries~\cite{Laha25}. Extensive searches have also been done from a plethora of other  sources such as X-ray binaries~\cite{IceCubeXraybinary}, Fast radio bursts~\cite{DesaiFRB}, pulsars~\cite{Pasumarti2}, pulsar wind nebulae~\cite{IceCubePWN}, red dwarfs~\cite{Shifa}, galaxy mergers~\cite{Lahamergers}, gamma-ray bursts~\cite{IceCubeGRB}, supernovae remnants~\cite{CCSN}, choked-jet supernovae~\cite{Chang24} and many more. 

Galaxy clusters are another possible source of high energy neutrinos.
Galaxy clusters are the most massive gravitationally collapsed objects in the universe with masses of $\sim~10^{14} M_{\odot}$~\cite{Vikhlininrev}.
They have been detected throughout the electromagnetic spectrum from radio waves to hard X-rays and have been shown to be excellent laboratories for Cosmology~\cite{Allen2011,Miyatake25}, galaxy evolution~\cite{Hennig}, as well as fundamental Physics~\cite{Desai18,Bohringer16}.  Observations of radio relics in clusters show that they contain a non-thermal population of high energy electrons~\cite{Brunetti2014}. 
Although the early results from Fermi-LAT produced only null results at GeV energies~\cite{Fermi2010}, recent results have found evidence for  GeV gamma-ray emission  from a number of individual clusters~\cite{Manna24,Harale,ShangLi} as well as from a stacked analysis of clusters~\cite{Mannastacked,Keshet}. It has been shown that the integrated gamma-ray flux from clusters in the redshift range $z\leq 5$ could contribute up to 100\% of the diffuse gamma-ray  flux observed by Fermi-LAT above 100 GeV~\cite{Dolagnature}. Finally we note that accretion shocks in nearby galaxy clusters such as Virgo have also been proposed as potential sources of   ultra-high cosmic rays detected by the Pierre Auger Observatory~\cite{Nuza}.  The Telescope Array Collaboration has also found evidence for the existence of a source of ultrahigh-energy cosmic ray events in the Perseus-Pisces supercluster~\cite{TA}. 
A recent review of high energy multi-messenger observations of galaxy clusters can be found in ~\cite{Dolag26}.

The production of neutrinos in galaxy clusters has been estimated in a number of works~\cite{Hussain,Murase,Fang16,Zandanel15,Dolag26}. In one of these models based on cosmological MHD simulations~\cite{Dolag26}, the maximum  muon neutrino energy flux  for the Virgo cluster 
has been estimated to be $E^2 \phi_{\nu} \sim  10^{-8} ~\rm{GeV~cm^{-2}~s^{-1}}$ for $E_{\nu}=0.1$ TeV,  (cf. Fig 4 in ~\cite{Dolag26}) after assuming cluster thermal energy $\sim 2 \times 10^{62}$ ergs, a power-law distribution of cosmic rays with spectral index $\sim 2.0$, and maximum cutoff energy equal to $10^{16}-10^{17}$ eV.

In a recent work, ~\citet{Harale}  (see also ~\cite{ShangLi}) found 4$\sigma$ evidence for gamma-ray emission from the galaxy cluster
Abell 119 in the energy range from 100 MeV to 1 TeV using 14 years of Fermi-LAT data. The cluster Abell 119 is one of the most massive clusters in the nearby universe, which is undergoing a merger~\cite{Hernan}. This cluster is located at a redshift $z=0.044$~\cite{Hernan} with RA = $14.0357^{\circ}$ and Dec. = $-1.20021^{\circ}$~\cite{Arnaud}  for the epoch J2000, with  a mass of $\sim 3 \times 10^{14} M_{\odot}$~\cite{Smith04}. Its  brightest cluster galaxy is a cD galaxy~\cite{Watson}, where a cD galaxy corresponds to a type-D elliptical galaxy, which is believed to grow via mergers of galaxies that spiral into the center of a galaxy cluster~\cite{Kormendy89}. Abell 119 is a dynamically complex cluster with evidence of multiple substructures with signatures of ongoing mergers.
Motivated by the recent  gamma-ray detection, a search  at MeV energies and hard X-rays from this cluster was carried out using the COMPTEL and INTEGRAL satellites, but no signal was  detected~\cite{Manna26}. 
This cluster has also been extensively imaged in X-rays using the Chandra satellite~\cite{Watson} (and references therein.)   It has also been imaged in radio waves at 1.4 GHz as part of the FIRST survey~\cite{Becker} and 150 MHz as part of the TIFR GMRT Sky Survey~\cite{Interna}. Abell 119 has also been detected in mm-wave  using the Sunyaev-Zeldovich effect by the Planck Satellite~\cite{Planck16}. Although no dedicated search for ultra-high cosmic rays from Abell 119 has been conducted, its distance—approximately ten times greater than that of the Virgo Cluster makes it unlikely to account for any of the events observed by the Pierre Auger Observatory, through accretion shocks via the mechanism proposed in ~\cite{Nuza}.

The gamma-ray signal detected in ~\cite{Harale} was found to be consistent with $\pi^{0}$ decay in the intra-cluster medium (ICM). These neutral pions are believed to be remnants of hadronic interactions of cosmic-ray protons in the ICM. The ratio of cosmic ray proton to thermal energy was estimated to be $\sim$ 8\% with a spectral slope of $2.2 \pm 0.3$~\cite{Harale}. However, despite the statistical significance of   $4\sigma$, \citet{Harale} cautioned that  there are still uncertainties due to source localization, PSF detector and other instrumental limitations, which must be resolved, before a definitive detection claim can be made. \rthis{Therefore,  at the time of writing the detection of gamma-rays from this cluster can only be considered as tentative. }
They argued that further confirmation for the hadronic origin of the gamma-ray emission can only be obtained through the detection of high energy neutrinos. The expected all-flavor differential neutrino energy flux was estimated to be $E^2\phi_{\nu} \approx 3 \times 10^{-10}$~GeV~$\rm{cm^{-2}~s^{-1}~sr^{-1}}$~\cite{Harale} at neutrino energy of 1 GeV (Harale, private communication), or the muon-neutrino flux is $\approx 10^{-10}$~GeV~$\rm{cm^{-2}~s^{-1}~sr^{-1}}$ at 1 GeV. In order to extrapolate this flux to high energies we assume a neutrino spectral index of 2.0, which \rthis{is within $1\sigma$} of the best-fit  gamma-ray spectral slope. Further, we also assume isotropic emission for  this model  and rescale the expected differential muon neutrino fluxes (provided in units of  GeV~$\rm{cm^{-2}~s^{-1}~sr^{-1}}$) by $4\pi$ to convert them into units of GeV~$\rm{cm^{-2} s^{-1}}$. 
If we rescale the muon neutrino flux estimated in \cite{Dolag26} according to the distance to Abell 119, we get $E^2\phi_{\nu} \sim  3 \times 10^{-10}$~GeV~$\rm{cm^{-2}~s^{-1}}$ at 0.1 TeV
which is half the flux  compared to the prediction in  ~\cite{Harale}, when extrapolated to 0.1 TeV. However, at higher energies the predicted flux in ~\cite{Dolag26} falls much more steeply compared to the estimate in \cite{Harale}. 
A comparison between the two theoretical models as a function of  energy is shown later on in this manuscript.

Other possibilities for the signal include Inverse-Compton scattering of Cosmic Microwave Background photons by cosmic ray electrons within the cluster~\cite{Blasi} or from point sources such as AGNs, star-forming galaxies or radio galaxies~\cite{Ackermann16}. However, based on the observed signal morphology, ~\citet{Harale} have argued that the point source model is ruled out, indicating that the emission  comes from the ICM, \rthis{(assuming the detection is confirmed)} and hence is likely to be hadronic. 
The gamma-ray flux  from Inverse-Compton scattering   has been estimated to be much smaller than that produced from hadronic emission for some other clusters~\cite{MannaMinot}.
We also note that  a gamma-ray signal could arise from dark matter annihilation~\cite{Maxim}, if the dark matter is a weakly interacting massive particle (WIMP). However, confirming whether the observed signal originates from WIMP annihilation would require a dedicated analysis using a template based on the expected annihilation-induced gamma-ray emission.

%The production of neutrinos in galaxy clusters has been  estimated  in a number of works~\cite{Hussain,Murase,Fang16,Zandanel15}. In one of these models~\cite{Hussain}, the differential neutrino energy flux  for a generic individual mass cluster $10^{14} M_{\odot}$ has been estimated  to range from  $\sim~10^{-8}-10^{-13}~\rm{GeV~cm^{-2}~s^{-1}~sr^{-1}}$ in the energy range from 100 to $10^{4}$ TeV, if the cosmic ray sources are located at the center of the cluster, and if  2\%  of the cluster luminosity is converted to cosmic rays (cf. Fig 12 in ~\cite{Hussain}). However, this flux can get reduced by three orders of magnitude if the CR sources are located at 1~Mpc from the center of the cluster. However, the aforementioned calculated flux is very much model-dependent and the calculated flux depends on the cosmological MHD simulations  used for the analysis. Other works have estimated lower values for the calculated flux.

%This work is a followup of our previous works where we looked for neutrino emission from FRBs~\cite{DesaiFRB}, pulsars~\cite{Pasumarti,Pasumarti2}, red dwarfs~\cite{Shifa} and magnetars~\cite{Shifa2026}. 
Motivated by this result, we search for neutrino emission from this galaxy cluster using the IceCube 10-year publicly available muon track data. 
The IceCube Collaboration has also previously reported null results in their searches for high energy neutrinos from individual nearby clusters~\cite{IceCube11,IceCube13} as well as from a stacked sample of SZ selected Planck clusters~\cite{IceCubeclusters}. They showed that galaxy clusters can contribute at most 4.6\% of the diffuse neutrino flux at 100 TeV~\cite{IceCubeclusters}.
A statistically significant detection of neutrinos from clusters would constitute evidence for proton acceleration~\cite{IceCubeclusters}. The limits obtained by the IceCube collaboration from the stacked analysis~\cite{IceCubeclusters} already rule out 
the MHD model of neutrino production within clusters proposed in ~\cite{Hussain}.
However, we are not aware of a previous search for neutrinos from Abell 119 from IceCube. Given the recent detection of gamma-rays from this cluster, it is timely to conduct a  search for neutrinos from this cluster. 

Therefore, we carry out a search for neutrino emission from this cluster using the 10-year muon track data which has been made publicly available by the IceCube collaboration. This work is a follow-up of our previous work in which we looked for neutrino emission from FRBs~\cite{DesaiFRB}, pulsars~\cite{Pasumarti,Pasumarti2}, red dwarfs~\cite{Shifa}, and magnetars~\cite{Shifa2026}. 

This manuscript is structured as follows. We describe the dataset used for the analysis in Sect.~\ref{sec:dataset}. The methodology is detailed in Sect.~\ref{sec:analysis}. Our results are presented in Sect.~\ref{sec:results}, followed by the conclusions in Sect.~\ref{sec:conclusions}.

\section{Dataset}
\label{sec:dataset}
For this analysis, we utilize neutrino events from the publicly available 10-year muon track dataset provided by the IceCube Collaboration~\cite{IceCubedata}.~\footnote{While this manuscript was in preparation, the IceCube collaboration had another data release encompassing 14 years of data~\cite{IceCube14}. For this work, we use the 10-year muon track dataset. \rthis{With the  14-year dataset we would expect the limits to scale with square root of total exposure, and hence should be close to that obtained in this work.}}  This dataset \rthis{ (referred to as IceTracks-DR1)} contains a total of 1,134,431 events recorded between April 2008 (IC-40 configuration) and July 2018 (IC86-VII), spanning multiple detector configurations with varying exposure times. For each event, the dataset includes the reconstructed right ascension (RA), declination (DEC), angular uncertainty, and muon energy.

In addition, we employ a modified version of this dataset presented in~\cite{Beacom}, in which duplicate events present in the original IceCube release have been removed. This corrected dataset is publicly accessible\footnote{Available at \url{https://github.com/beizhouphys/IceCube_data_2008--2018_double_counting_corrected}}.

\section{Analysis}
\label{sec:analysis}
The analysis employs the unbinned maximum likelihood ratio method, following recent works (outside the IceCube Collaboration)~\cite{Kamionkowski,Li22,Smith21}, which are based on the framework originally proposed in~\cite{Montaruli}.

We select neutrino events with declinations lying within $\pm 5^\circ$ of the galaxy cluster.\footnote{We also did a similar analysis within $\pm 3^{\circ}$, but do not report these results as they were almost the same as those presented here.}
For a dataset containing $N$ events, of which $n_s$ are assumed to originate from the source, the probability density function (PDF) for the $i$-th event is given by
\begin{equation}
    P_i = \frac{n_s}{N} S_i + \left(1 - \frac{n_s}{N}\right) B_i,
\label{eq:1}
\end{equation}
where $S_i$ and $B_i$ denote the signal and background PDFs, respectively.

The signal PDF is modeled as a two-dimensional Gaussian:
\begin{equation}
    S_i = \frac{1}{2\pi \sigma_i^2} \exp\left(-\frac{|\theta_i - \theta_s|^2}{2\sigma_i^2}\right),
    \label{eq:L}
\end{equation}
where $|\theta_i - \theta_s|$ represents the angular separation between the reconstructed neutrino direction and the source position, and $\sigma_i$ is the angular uncertainty associated with the event. \rthis{Note that Eq.~\ref{eq:L} assumes that Abell 119 is a point source. The angular radius of this cluster parameterized by $\theta_{200}=R_{200}/D_A$ is equal to $0.6^{\circ}$, where $D_A$ is the angular diameter distance to the cluster and $R_{200}$ is the radius corresponding to an overdensity of 200 compared to the critical density of the universe. This is comparable to the median uncertainty ($0.59^{\circ}$) in the reconstructed direction around the declination of the cluster.  Therefore, we do the analysis assuming a point source. In Appendix A, we shall do an extended source analysis taking into account the angular extent of the cluster.}

The likelihood function for the full dataset is then constructed as follows.
\begin{equation}
    \mathcal{L}(n_s) = \prod_{i=1}^{N} P_i.
\end{equation}

We assume that the background events are uniformly distributed in right ascension within the selected declination band. Under this assumption, the background PDF is given by
\begin{equation}
    B_i = \frac{1}{\Omega_{\delta \pm 5^\circ}},
\end{equation}
where $\Omega_{\delta \pm 5^\circ}$ denotes the solid angle corresponding to the declination band $\delta \pm 5^\circ$ around the source. \rthis{A plot of this  background PDF in the $\pm 5^{\circ}$ region of interest in bins of $1^{\circ}$ can be found in Fig.~\ref{fig:bg}. We note that at the declination of the cluster, the detector response varies rapidly.}

To quantify the presence of a signal, we define the test statistic (TS) as
\begin{equation}
    TS = 2 \log \frac{\mathcal{L}(\hat{n}_s)}{\mathcal{L}(0)},
\label{eq:ts}
\end{equation}
where $\hat{n}_s$ is the value of $n_s$ that maximizes the likelihood function.

Under the null hypothesis, the TS asymptotically follows a half-$\chi^2$ distribution with one degree of freedom~\cite{Mattox96,Cowan11}.  In a previous work, we have also independently confirmed  (based on random sky locations  in RA and $\sin (\delta)$ ) that the TS statistic for the IceCube data obeys  a half-$\chi^2$ distribution for the null hypothesis~\cite{Shifa2026}.
%However, we note that randomizing in both RA and $\sin \delta$ is a simplification, since the detector background rate is zenith/declination-dependent and hence  spatial clustering could  affect the TS distribution under the background hypothesis. 
We also note that the  IceCube collaboration has also independently verified that TS distribution obeys a half $\chi^2$ distribution for one degree of freedom~\cite{IceCubeDM}, and therefore we assume the same for calculating upper limits. 
The statistical significance of a potential signal is given by $\sqrt{TS}$~\cite{Mattox96}. Consequently, a $5\sigma$ detection corresponds to $TS > 25$.

\section{Results}
\label{sec:results}

We obtain a best-fit test statistic value of $TS = 0.47$ for Abell 119, corresponding to $\hat{n}_s = 7.5$. The TS value is much smaller than 25, which is needed to claim a detection, and hence  is consistent with a background fluctuation. Therefore, we conclude that there is no detection of neutrinos from this cluster and hence set upper limits.

To derive the 90\% confidence level (c.l.) upper limit on the number of signal events, we determine the value of $n_s$ for which the test statistic decreases by 1.64 from its maximum value, i.e., $\mathrm{TS}(n_s) = \mathrm{TS}_{\max} - 1.64$~\cite{IceCubeDM}. This yields an upper limit of $n_s^{90} = 23.5$.

\begin{figure}
\centering
\includegraphics[width=0.8\columnwidth]{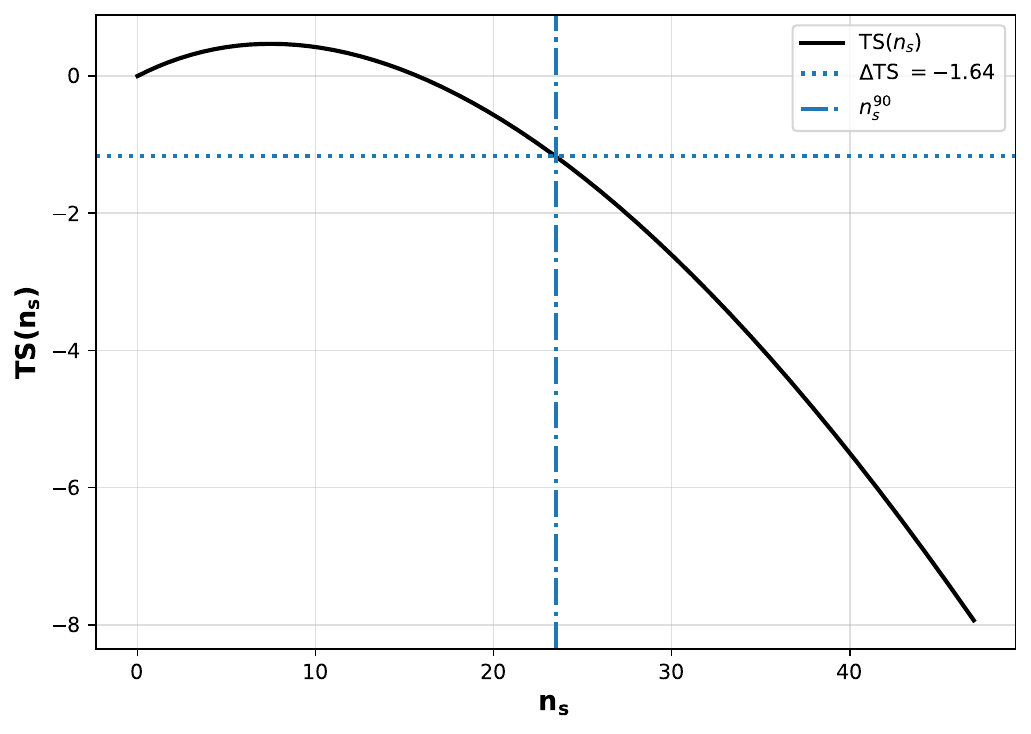}
\caption{This plot shows TS $(n_s) =2 \log \frac{\mathcal{L} (n_s)}{\mathcal{L}(0)}$ as a function of $n_s$. The value of $\hat{n}_s$ that maximizes the likelihood function is equal to  7.5 and the 90\% upper limit corresponding to $\Delta \mathrm{TS}$=-1.64 occurs at $n_s=23.5$. \rthis{Here, $\Delta \mathrm{TS} = \mathrm{TS} (n_s)- \mathrm{TS} (\hat{n}_s)$}.}
\label{fig:PL}
\end{figure}

\begin{figure}
\centering
\includegraphics[width=0.8\columnwidth]{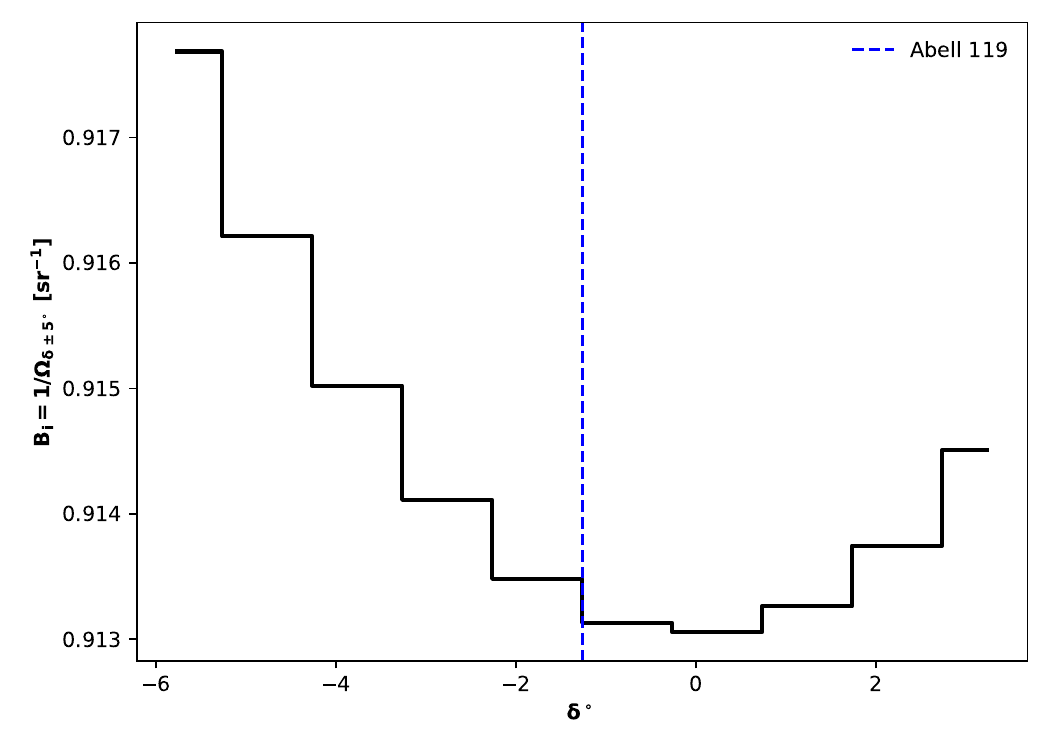}
\caption{\rthis{Background PDF in region of interest corresponding to a declination band of $\pm 5^{\circ}$ around the source, Abell 119. The blue dashed line represents the declination of Abell 119. }}
\label{fig:bg}
\end{figure}

To translate this into a limit on the differential neutrino flux, we follow the procedure outlined in~\cite{Shifa2026}. We assume a power-law form for the neutrino spectrum:
\begin{equation}
    \phi(E) = \phi_0 \left( \frac{E}{E_0} \right)^{-\Gamma},
    \label{eq:phinu}
\end{equation}
where $\phi(E)$ is the muon neutrino flux, $\phi_0$ is the normalization at the pivot energy $E_0$, and $\Gamma$ is the spectral index. The pivot energy is taken to be $E_0 = 10^5~\mathrm{GeV}$, and $\phi(E)$ is expressed in units of $\mathrm{GeV}^{-1}~\mathrm{cm}^{-2}~\mathrm{s}^{-1}$.

The total expected  number of signal events ($n_s$) is then computed by adding over all the detector seasons:
\begin{equation}
n_s = \sum_{k} T_k \int A_{\mathrm{eff}}^k (E_{\nu}, \delta) \, \phi(E_{\nu}) \, dE_{\nu},
\label{eq:acc}
\end{equation}
where $T_k$ denotes the livetime of the $k$-th season, and $A_{\mathrm{eff}}^k (E_{\nu}, \delta)$ is the IceCube effective area in the $k$-th season, which depends on both neutrino energy and source declination. We have used ten seasons  to calculate the exposure: \texttt{IC40}, \texttt{IC59}, \texttt{IC79}, \texttt{IC86\_I}, \texttt{IC86\_II}, \texttt{IC86\_III}, \texttt{IC86\_IV}, \texttt{IC86\_V}, \texttt{IC86\_VI}, \texttt{IC86\_VII}. We note that the effective area includes contributions from charged current and neutral current interactions~\cite{IceCube13}, and does not take into account contributions from the  subdominant but important high energy neutrino interaction channels, such as neutrino-nucleus W-boson production and  final-state radiation.
After plugging in the expression for $\phi(E_{\nu})$ from Eq.~\eqref{eq:phinu}, we obtain:
\begin{equation}
n_s = \phi_0 \sum_{k} T_k \int A_{\mathrm{eff}}^k (E_{\nu}, \delta) \,  \left( \frac{E_{\nu}}{E_0} \right)^{-\Gamma} \, dE_{\nu},
\label{eq:acc2}
\end{equation}
In evaluating this expression, we adopt a spectral index $\Gamma = 2.0$, \rthis{as this value is within $1\sigma$ of the best-fit spectral index obtained in ~\cite{Harale}}. The integrand in Eq.~\eqref{eq:acc} ranges over all the energy bins for which effective area is non-zero. This lower limit is around 100 GeV and the upper limit around $8 \times 10^6$ GeV. \rthis{These limits are determined by the lower and upper energy bounds for which the IceCube collaboration provides 
effective-area data.}

The  muon neutrino flux upper limit ($\phi_0$) at $E_{\nu} = 100~\mathrm{TeV}$ is then obtained by dividing the upper limit on the number of signal events ($n_s^{90}$) by  the  RHS of Eq.~\eqref{eq:acc2}. This yields
\[
\phi_0 = 1.2 \times 10^{-19}~\mathrm{GeV}^{-1}~\mathrm{cm}^{-2}~\mathrm{s}^{-1}.
\]
To convert this to a differential muon neutrino flux limit, we multiply by $E^2$ and obtain the following value for $E^2 \phi_{\nu}$ at the pivot energy of 100 TeV:
\[
E^2 \phi_{\nu} = 1.2 \times 10^{-9}~\mathrm{GeV}~\mathrm{cm}^{-2}~\mathrm{s}^{-1}
\]

The differential muon neutrino flux ($E^2 \phi_{\nu} (E)$) as a function of neutrino energy for the spectral index of $\Gamma=2.0$ can be found in Fig.~\ref{fig:sens}. We also compare our 90\% c.l. upper limit with the theoretical models in ~\cite{Harale} and \cite{Dolag26}, after suitable rescalings as explained in Section~\ref{sec:intro}. 
We find that  our 90\% c.l. flux limit is about one to two orders  of magnitude  larger than both the theoretical models~\cite{Harale,Dolag26}. Therefore, our limit cannot  rule out either of the two theoretical models considered.

We now compare our upper limit with the 90\% c.l. median sensitivity for $E^2 \phi_{\nu}$,  derived   using 10 years of IceCube data also assuming  an $E^{-2}$ spectrum (cf. Fig.~3 of \cite{IceCube20}). 
After extrapolating to the declination of this cluster, the expected sensitivity of IceCube was found to be  $ \sim 2.7\times10^{-10}$ GeV cm$^{-2}$s$^{-1}$. The corresponding value which we obtain for this spectral index is $\sim 1.2 \times 10^{-9}$ GeV cm$^{-2}$s$^{-1}$.
Therefore, the projected IceCube sensitivity around the declination of this cluster is about four times lower than our 90\% c.l. upper limit. \rthis{One reason  why our limits are higher than the projected IceCube sensitivity  is because the IceCube 10-year  point source analysis~\cite{IceCube20} also accounts for the neutrino energy term in the unbinned maximum likelihood, unlike our analysis. } 

In Fig.~\ref{fig:sens}, we also compare our results to the projected sensitivities of future neutrino detectors from this cluster assuming  a livetime of 10 years. For this purpose, we used either the  published median sensitivity (from these detectors) at 90\% c.l. or the 5$\sigma$ discovery potential (in case the 90\% c.l. sensitivity is not available)  at the same declination as Abell 119. 
The median sensitivity is typically a factor of 2-3 below the $5\sigma$ discovery potential, so both the quantities should provide reasonable estimates of the expected reach. Since all the projected sensitivities from the future detectors were  estimated for a spectral index of 2.0, we also compare these results with our limit assuming the same spectral index.
 Our  limit, superposed along  with the projected sensitivities / discovery potential of future detectors such as IceCube-Gen2~\cite{Gen2}, KM3NeT/ACRA230~\cite{2024APh...16202990A}, P-ONE~\cite{2022icrc.confE..24R} and TRIDENT~\cite{2023NatAs...7.1497Y} can be found in Fig.~\ref{fig:sens}.
Note that  for this comparison, we did not include HUNT~\cite{HUNT}, since it has a  very high energy threshold compared to IceCube.  Among all these detectors, only TRIDENT \rthis{and IceCube-Gen2} can rule out (or confirm) the prediction for the hadronic emission model~\cite{Harale}. We note however  the flux prediction for the neutrino production  mechanism model in \cite{Dolag26} at  the distance to Abell 119  can only be probed by TRIDENT in the energy range up to 1 TeV. 
\rthis{Among all detectors, TRIDENT has the largest effective area at the declination of Abell 119,  and hence has the highest sensitivity. For comparison, TRIDENT  effective area is  $\sim 8\times10^2 \mathrm{m}^2$~\cite{2023NatAs...7.1497Y} compared to $10 \mathrm{m}^2$ for IceCube-Gen2~\cite{Gen2} and $2\times10 \mathrm{m}^2$ for KM3NeT/ARCA230~\cite{2025Natur.638..376K} at 100 TeV at the  declination of the Abell 119 cluster.}

\iffalse
\begin{figure}
\centering
\includegraphics[width=\linewidth]{sensitivity_gamma22_div3_virgo_final.pdf}
    \caption{ 
    Integrated  muon neutrino flux limits for IceCube    as a function of energy 
    assuming a spectral index $\Gamma=2.0$, along with theoretical predictions from ~\cite{Harale} (assuming a spectral index of 2.0) and \cite{Dolag26} (after rescaling according to the distance to Abell 119). We find that our limits cannot rule out either of the two theoretical models.}
    \label{fig:gamma22}
\end{figure}
\fi

\begin{figure}
\centering
 \includegraphics[width=\linewidth]{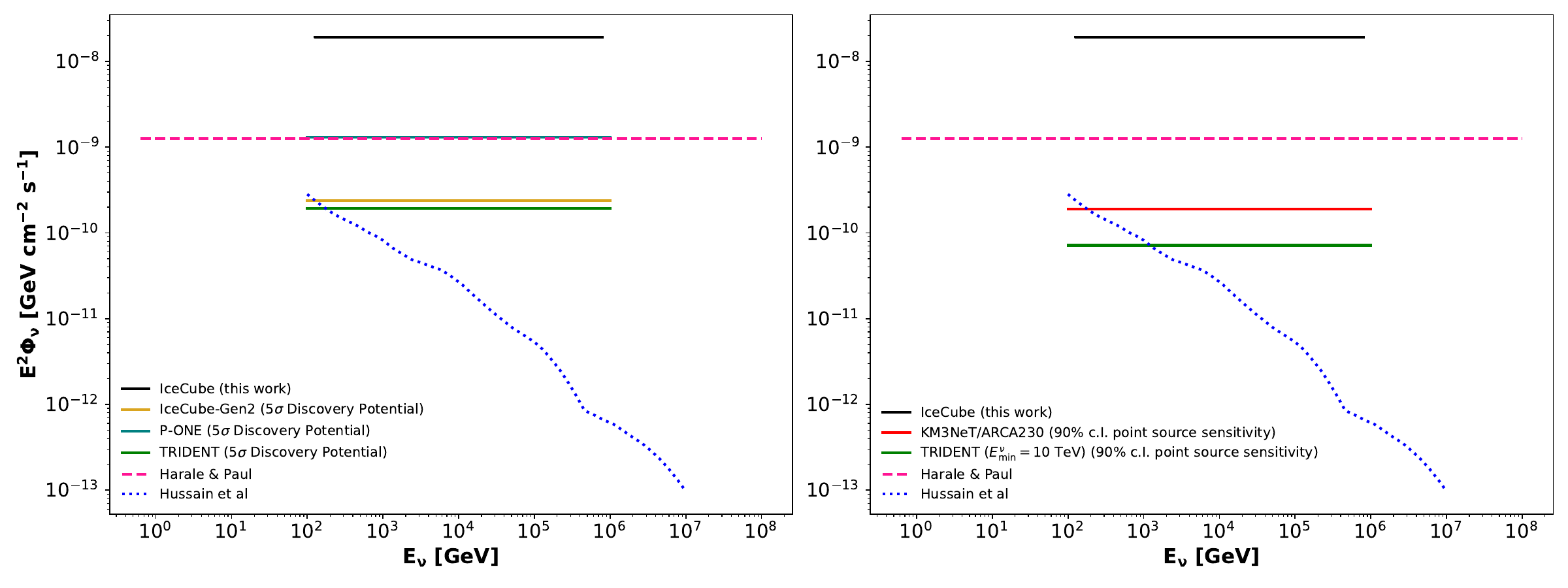}
    \caption{ Differential muon neutrino flux limit for IceCube as a function of neutrino energy 
    assuming a spectral index $\Gamma=2$,
    along with theoretical predictions from~\cite{Harale} (for a spectral index of 2.0)   and~\cite{Dolag26}. 
    The left panel contains  5$\sigma$ discovery potential for      IceCube-Gen2~\cite{2022icrc.confE1184O},  P-ONE~\cite{2022icrc.confE..24R}, and TRIDENT~\cite{2023NatAs...7.1497Y}, while the right panel contains the      90\% c.l. median sensitivity  estimates for 
    KM3Net/ARCA~\cite{2024EPJC...84..885K} and TRIDENT~\cite{2023NatAs...7.1497Y}, assuming 10 years of exposure.  All the sensitivity estimates for future detectors  have also assumed a spectral index of 2.0. }
    \label{fig:sens}
\end{figure}

\section{Conclusions}
\label{sec:conclusions}
In this work, we search for high energy muon neutrino emission from the galaxy cluster Abell 119, motivated by a  recent \rthis{tentative} detection of gamma-rays from this cluster, which was attributed to hadronic processes~\cite{Harale}. This work  made a prediction, that if the gamma-ray emission was due to hadronic processes, one would expect muon neutrino flux at 1 GeV to be $E^2\phi_{\nu} \approx 1.25 \times 10^{-9}$ GeV~$\rm{cm^{-2}} s^{-1}$\cite{Harale}, after assuming isotropic emission.

Although a detailed stacked analysis of neutrino emission from galaxy clusters has been done by the IceCube Collaboration~\cite{IceCubeclusters}, no previous limit has been obtained from Abell 119.
For this purpose, we performed  a time-integrated search using the 10-year muon track data with the unbinned maximum likelihood method~\cite{Montaruli}. The value of the test statistic  used to ascertain the presence of a signal was found to be $\sim 0.5$, which implies that there is no statistically significant evidence for neutrino emission from this cluster. We therefore calculated the muon  neutrino flux limit from this cluster at 90\% c.l., assuming a spectral index of  2.0. This plot can be found in Fig.~\ref{fig:sens}. 
At a pivot energy of 100 TeV, the corresponding upper limit is  $1.2 \times 10^{-9}~\mathrm{GeV}~\mathrm{cm}^{-2}~\mathrm{s}^{-1}$, assuming a spectral index of 2.0.
%SD: change the above number
The expected neutrino flux from the hadronic emission model in ~\cite{Harale} is about an order of magnitude below our IceCube based neutrino flux limit and hence cannot be ruled out. The flux prediction from the MHD model proposed in ~\cite{Dolag26} is about two orders of magnitude lower than our upper limit. 
%\rthis{The differential neutrino flux as a function of neutrino energy can be found in Fig.~\ref{fig:gamma22}.}

%Although this is of the same order of magnitude as the estimate in \cite{Harale}, our limit has been calculated at 100 TeV, whereas the prediction in ~\cite{Harale} was made at 1 GeV, which precludes a direct comparison. We also compared our limit with another theoretical model for neutrino production in the Perseus cluster, based on cosmological MHD simulations,  and find that the estimate for the corresponding emission from Abell 119 cluster after rescaling based on distance, is about one order of magnitude lower than our 90\% c.l. upper limit.

We also compare the  predictions for this model with projected 90\% c.l. sensitivity estimates  or $5\sigma$ discovery potential for  a few next generation neutrino detectors assuming a livetime of 10 years. Among the next generation detectors, only TRIDENT \rthis{and IceCube-Gen2} have the expected sensitivity to test  the hadronic interpretation proposed  for the  tentative gamma-ray signal observed in Abell 119. 
If neither detector  observes a  signal from this cluster, while the Fermi-LAT  signal is confirmed by next generation gamma-ray telescopes such as Cherenkov Telescope Array~\cite{CTA}, this would suggest  either a leptonic origin of  the signal or emission  from  nearby point sources.

Therefore, although galaxy clusters have already been ruled out as the dominant contributor to the diffuse astrophysical neutrino background observed by IceCube~\cite{IceCubeclusters}, neutrino flux limits from individual clusters such as Abell 119 can be used to constrain models of hadronic gamma-ray emission within clusters, demonstrating the synergy of neutrino astrophysics with gamma-ray astrophysics.

\begin{acknowledgments}
 We also thank Gajanan Harale, Saqib Hussain, and Fuyudi Zhang for helpful correspondence. \rthis{We also thank the anonymous referees for several useful  comments and feedback on the manuscript.}
\end{acknowledgments}

\bibliography{apssamp}

@ARTICLE{Harale,
       author = {{Harale}, Gajanan D. and {Paul}, Surajit},
        title = "{Excess of diffuse gamma-ray emission detected from the galaxy cluster Abell 119 from 14-year Fermi-LAT data}",
      journal = {\prd},
         year = 2025,
        month = nov,
       volume = {112},
       number = {10},
          eid = {103025},
        pages = {103025},
          doi = {10.1103/gn1q-pzx3},
archivePrefix = {arXiv},
       eprint = {2511.15559},
 primaryClass = {astro-ph.HE},
       adsurl = {https://ui.adsabs.harvard.edu/abs/2025PhRvD.112j3025H}
}

@ARTICLE{ShangLi,
       author = {{Li}, Shang and {Han}, Feng},
        title = "{Search for {\ensuremath{\gamma}}-Ray Emission from Cluster of Galaxies with Fermi-LAT Data}",
      journal = {\apj},
         year = 2026,
        month = feb,
       volume = {997},
       number = {2},
          eid = {227},
        pages = {227},
          doi = {10.3847/1538-4357/ae2bda},
       adsurl = {https://ui.adsabs.harvard.edu/abs/2026ApJ...997..227L}
}

@ARTICLE{Smith04,
       author = {{Smith}, Russell J. and {Hudson}, Michael J. and {Nelan}, Jenica E. and {Moore}, Stephen A.~W. and {Quinney}, Stephen J. and {Wegner}, Gary A. and {Lucey}, John R. and {Davies}, Roger L. and {Malecki}, Justin J. and {Schade}, David and {Suntzeff}, Nicholas B.},
        title = "{NOAO Fundamental Plane Survey. I. Survey Design, Redshifts, and Velocity Dispersion Data}",
      journal = {\aj},
         year = 2004,
        month = oct,
       volume = {128},
       number = {4},
        pages = {1558-1569},
          doi = {10.1086/423915},
       adsurl = {https://ui.adsabs.harvard.edu/abs/2004AJ....128.1558S}
}

@ARTICLE{IceCube14,
       author = {{Abbasi}, R. and {Ackermann}, M. and {Adams}, J. and {Aguilar}, J.~A. and {Ahlers}, M. and {Alameddine}, J.~M. and {Ali}, S. and {Amin}, N.~M. and {Andeen}, K. and {Arg{\"u}elles}, C. and {Ashida}, Y. and {Athanasiadou}, S. and {Axani}, S.~N. and {Babu}, R. and {Bai}, X. and {Balagopal V.}, A. and {Barwick}, S.~W. and {Basu}, V. and {Bay}, R. and {Beatty}, J.~J. and {Becker Tjus}, J. and {Behrens}, P. and {Beise}, J. and {Bellenghi}, C. and {Benkel}, S. and {BenZvi}, S. and {Berley}, D. and {Bernardini}, E. and {Besson}, D.~Z. and {Blaufuss}, E. and {Bloom}, L. and {Blot}, S. and {Bontempo}, F. and {Book Motzkin}, J.~Y. and {Boscolo Meneguolo}, C. and {B{\"o}ser}, S. and {Botner}, O. and {B{\"o}ttcher}, J. and {Braun}, J. and {Brinson}, B. and {Brisson-Tsavoussis}, Z. and {Burley}, R.~T. and {Butterfield}, D. and {Carloni}, K. and {Carpio}, J. and {Chau}, N. and {Chen}, Y.~C. and {Chen}, Z. and {Chirkin}, D. and {Choi}, S. and {Chubarov}, A. and {Clark}, B.~A. and {Collin}, G.~H. and {Coloma Borja}, D.~A. and {Connolly}, A. and {Conrad}, J.~M. and {Cowen}, D.~F. and {De Clercq}, C. and {DeLaunay}, J.~J. and {Delgado}, D. and {Delmeulle}, T. and {Deng}, S. and {Desiati}, P. and {de Vries}, K.~D. and {de Wasseige}, G. and {DeYoung}, T. and {D{\'\i}az-V{\'e}lez}, J.~C. and {DiKerby}, S. and {Ding}, T. and {Dittmer}, M. and {Domi}, A. and {Draper}, L. and {Dueser}, L. and {Durnford}, D. and {Dutta}, K. and {DuVernois}, M.~A. and {Ehrhardt}, T. and {Eidenschink}, L. and {Eimer}, A. and {Eldridge}, C. and {Eller}, P. and {Ellinger}, E. and {Els{\"a}sser}, D. and {Engel}, R. and {Erpenbeck}, H. and {Esmail}, W. and {Eulig}, S. and {Evans}, J. and {Evenson}, P.~A. and {Fan}, K.~L. and {Fang}, K. and {Farrag}, K. and {Fazely}, A.~R. and {Fedynitch}, A. and {Feigl}, N. and {Finley}, C. and {Fox}, D. and {Franckowiak}, A. and {Fukami}, S. and {F{\"u}rst}, P. and {Gallagher}, J. and {Ganster}, E. and {Garcia}, A. and {Garcia}, M. and {Genton}, E. and {Gerhardt}, L. and {Ghadimi}, A. and {Glaser}, C. and {Gl{\"u}senkamp}, T. and {Gonzalez}, J.~G. and {Goswami}, S. and {Granados}, A. and {Grant}, D. and {Gray}, S.~J. and {Griffin}, S. and {Groth}, K.~M. and {Guevel}, D. and {G{\"u}nther}, C. and {Gutjahr}, P. and {Ha}, C. and {Hallgren}, A. and {Halve}, L. and {Halzen}, F. and {Hamacher}, L. and {Handt}, M. and {Hanson}, K. and {Hardin}, J. and {Harnisch}, A.~A. and {Hatch}, P. and {Haungs}, A. and {H{\"a}u{\ss}ler}, J. and {Helbing}, K. and {Hellrung}, J. and {Henke}, B. and {Hennig}, L. and {Henningsen}, F. and {Heuermann}, L. and {Hewett}, R. and {Heyer}, N. and {Hickford}, S. and {Hidvegi}, A. and {Hill}, C. and {Hill}, G.~C. and {Hmaid}, R. and {Hoffman}, K.~D. and {Hollnagel}, A. and {Hooper}, D. and {Hori}, S. and {Hoshina}, K. and {Hostert}, M. and {Hou}, W. and {Hrywniak}, M. and {Huber}, T. and {Hultqvist}, K. and {Hymon}, K. and {Ishihara}, A. and {Iwakiri}, W. and {Jacquart}, M. and {Jain}, S. and {Janik}, O. and {Jansson}, M. and {Jin}, M. and {Kamp}, N. and {Kang}, D. and {Kang}, W. and {Kappes}, A. and {Kardum}, L. and {Karg}, T. and {Karle}, A. and {Katil}, A. and {Kauer}, M. and {Kelley}, J.~L. and {Khanal}, M. and {Khatee Zathul}, A. and {Kheirandish}, A. and {Kim}, T. and {Kimku}, H. and {Kirchner}, F. and {Kiryluk}, J. and {Klein}, C. and {Klein}, S.~R. and {Kobayashi}, Y. and {Koch}, S. and {Kochocki}, A. and {Koirala}, R. and {Kolanoski}, H. and {Kontrimas}, T. and {K{\"o}pke}, L. and {Kopper}, C. and {Koskinen}, D.~J. and {Koundal}, P. and {Kowalski}, M. and {Kozynets}, T. and {Kravka}, A. and {Krieger}, N. and {Krishnan}, T. and {Kruiswijk}, K. and {Krupczak}, E. and {Kumar}, A. and {Kun}, E.},
        title = "{IceCube Second Track Data Release IceTracks-DR2: Data from 2008-2022 for Neutrino Source Searches}",
      journal = {arXiv e-prints},
         year = 2026,
        month = may,
          eid = {arXiv:2605.19040},
        pages = {arXiv:2605.19040},
          doi = {10.48550/arXiv.2605.19040},
archivePrefix = {arXiv},
       eprint = {2605.19040},
 primaryClass = {astro-ph.HE},
       adsurl = {https://ui.adsabs.harvard.edu/abs/2026arXiv260519040A}
}

@ARTICLE{Pasumarti,
       author = {{Pasumarti}, Vibhavasu and {Desai}, Shantanu},
        title = "{Search for spatial coincidence between IceCube neutrinos and radio pulsars}",
      journal = {\jcap},
         year = 2022,
        month = dec,
       volume = {2022},
       number = {12},
          eid = {002},
        pages = {002},
          doi = {10.1088/1475-7516/2022/12/002},
archivePrefix = {arXiv},
       eprint = {2210.12804},
 primaryClass = {astro-ph.HE},
       adsurl = {https://ui.adsabs.harvard.edu/abs/2022JCAP...12..002P}
}

@ARTICLE{Lahamergers,
       author = {{Bouri}, Subhadip and {Parashari}, Priyank and {Das}, Mousumi and {Laha}, Ranjan},
        title = "{First search for high-energy neutrino emission from galaxy mergers}",
      journal = {\prd},
         year = 2025,
        month = mar,
       volume = {111},
       number = {6},
          eid = {063059},
        pages = {063059},
          doi = {10.1103/PhysRevD.111.063059},
archivePrefix = {arXiv},
       eprint = {2404.06539},
 primaryClass = {astro-ph.HE},
       adsurl = {https://ui.adsabs.harvard.edu/abs/2025PhRvD.111f3059B}
}

@ARTICLE{IcecubeXraybinary,
       author = {{Abbasi}, R. and {Ackermann}, M. and {Adams}, J. and {Aguilar}, J.~A. and {Ahlers}, M. and {Ahrens}, M. and {Alameddine}, J.~M. and {Alves}, A.~A. and {Amin}, N.~M. and {Andeen}, K. and {Anderson}, T. and {Anton}, G. and {Arg{\"u}elles}, C. and {Ashida}, Y. and {Axani}, S. and {Bai}, X. and {Balagopal}, V.~A. and {Barwick}, S.~W. and {Bastian}, B. and {Basu}, V. and {Baur}, S. and {Bay}, R. and {Beatty}, J.~J. and {Becker}, K.-H. and {Tjus}, J. Becker and {Beise}, J. and {Bellenghi}, C. and {Benda}, S. and {BenZvi}, S. and {Berley}, D. and {Bernardini}, E. and {Besson}, D.~Z. and {Binder}, G. and {Bindig}, D. and {Blaufuss}, E. and {Blot}, S. and {Boddenberg}, M. and {Bontempo}, F. and {Borowka}, J. and {B{\"o}ser}, S. and {Botner}, O. and {B{\"o}ttcher}, J. and {Bourbeau}, E. and {Bradascio}, F. and {Braun}, J. and {Brinson}, B. and {Bron}, S. and {Brostean-Kaiser}, J. and {Browne}, S. and {Burgman}, A. and {Burley}, R.~T. and {Busse}, R.~S. and {Campana}, M.~A. and {Carnie-Bronca}, E.~G. and {Chen}, C. and {Chen}, Z. and {Chirkin}, D. and {Choi}, K. and {Clark}, B.~A. and {Clark}, K. and {Classen}, L. and {Coleman}, A. and {Collin}, G.~H. and {Conrad}, J.~M. and {Coppin}, P. and {Correa}, P. and {Cowen}, D.~F. and {Cross}, R. and {Dappen}, C. and {Dave}, P. and {De Clercq}, C. and {DeLaunay}, J.~J. and {Delgado L{\'o}pez}, D. and {Dembinski}, H. and {Deoskar}, K. and {Desai}, A. and {Desiati}, P. and {de Vries}, K.~D. and {de Wasseige}, G. and {de With}, M. and {DeYoung}, T. and {Diaz}, A. and {D{\'\i}az-V{\'e}lez}, J.~C. and {Dittmer}, M. and {Dujmovic}, H. and {Dunkman}, M. and {DuVernois}, M.~A. and {Ehrhardt}, T. and {Eller}, P. and {Engel}, R. and {Erpenbeck}, H. and {Evans}, J. and {Evenson}, P.~A. and {Fan}, K.~L. and {Fazely}, A.~R. and {Fedynitch}, A. and {Feigl}, N. and {Fiedlschuster}, S. and {Fienberg}, A.~T. and {Finley}, C. and {Fischer}, L. and {Fox}, D. and {Franckowiak}, A. and {Friedman}, E. and {Fritz}, A. and {F{\"u}rst}, P. and {Gaisser}, T.~K. and {Gallagher}, J. and {Ganster}, E. and {Garcia}, A. and {Garrappa}, S. and {Gerhardt}, L. and {Ghadimi}, A. and {Glaser}, C. and {Glauch}, T. and {Gl{\"u}senkamp}, T. and {Gonzalez}, J.~G. and {Goswami}, S. and {Grant}, D. and {Gr{\'e}goire}, T. and {Griswold}, S. and {G{\"u}nther}, C. and {Gutjahr}, P. and {Haack}, C. and {Hallgren}, A. and {Halliday}, R. and {Halve}, L. and {Halzen}, F. and {Minh}, M. Ha and {Hanson}, K. and {Hardin}, J. and {Harnisch}, A.~A. and {Haungs}, A. and {Hebecker}, D. and {Helbing}, K. and {Henningsen}, F. and {Hettinger}, E.~C. and {Hickford}, S. and {Hignight}, J. and {Hill}, C. and {Hill}, G.~C. and {Hoffman}, K.~D. and {Hoffmann}, R. and {Hoshina}, K. and {Huang}, F. and {Huber}, M. and {Huber}, T. and {Hultqvist}, K. and {H{\"u}nnefeld}, M. and {Hussain}, R. and {Hymon}, K. and {In}, S. and {Iovine}, N. and {Ishihara}, A. and {Jansson}, M. and {Japaridze}, G.~S. and {Jeong}, M. and {Jin}, M. and {Jones}, B.~J.~P. and {Kang}, D. and {Kang}, W. and {Kang}, X. and {Kappes}, A. and {Kappesser}, D. and {Kardum}, L. and {Karg}, T. and {Karl}, M. and {Karle}, A. and {Katz}, U. and {Kauer}, M. and {Kellermann}, M. and {Kelley}, J.~L. and {Kheirandish}, A. and {Kin}, K. and {Kintscher}, T. and {Kiryluk}, J. and {Klein}, S.~R. and {Kochocki}, A. and {Koirala}, R. and {Kolanoski}, H. and {Kontrimas}, T. and {K{\"o}pke}, L. and {Kopper}, C. and {Kopper}, S. and {Koskinen}, D.~J. and {Koundal}, P. and {Kovacevich}, M. and {Kowalski}, M. and {Kozynets}, T. and {Krupczak}, E. and {Kun}, E. and {Kurahashi}, N. and {Lad}, N. and {Gualda}, C. Lagunas and {Lanfranchi}, J.~L. and {Larson}, M.~J. and {Lauber}, F. and {Lazar}, J.~P. and {Lee}, J.~W. and {Leonard}, K.},
        title = "{Search for High-energy Neutrino Emission from Galactic X-Ray Binaries with IceCube}",
      journal = {\apjl},
         year = 2022,
        month = may,
       volume = {930},
       number = {2},
          eid = {L24},
        pages = {L24},
          doi = {10.3847/2041-8213/ac67d8},
archivePrefix = {arXiv},
       eprint = {2202.11722},
 primaryClass = {astro-ph.HE},
       adsurl = {https://ui.adsabs.harvard.edu/abs/2022ApJ...930L..24A}
}

@ARTICLE{IceCubeGRB,
       author = {{Abbasi}, R. and {Ackermann}, M. and {Adams}, J. and {Agarwalla}, S.~K. and {Aguilar}, J.~A. and {Ahlers}, M. and {Alameddine}, J.~M. and {Amin}, N.~M. and {Andeen}, K. and {Anton}, G. and {Arg{\"u}elles}, C. and {Ashida}, Y. and {Athanasiadou}, S. and {Ausborm}, L. and {Axani}, S.~N. and {Bai}, X. and {A. Balagopal}, V. and {Baricevic}, M. and {Barwick}, S.~W. and {Basu}, V. and {Bay}, R. and {Beatty}, J.~J. and {Becker Tjus}, J. and {Beise}, J. and {Bellenghi}, C. and {Benning}, C. and {BenZvi}, S. and {Berley}, D. and {Bernardini}, E. and {Besson}, D.~Z. and {Blaufuss}, E. and {Blot}, S. and {Bontempo}, F. and {Book}, J.~Y. and {Boscolo Meneguolo}, C. and {B{\"o}ser}, S. and {Botner}, O. and {B{\"o}ttcher}, J. and {Braun}, J. and {Brinson}, B. and {Brostean-Kaiser}, J. and {Brusa}, L. and {Burley}, R.~T. and {Busse}, R.~S. and {Butterfield}, D. and {Campana}, M.~A. and {Carloni}, K. and {Carnie-Bronca}, E.~G. and {Chattopadhyay}, S. and {Chau}, N. and {Chen}, C. and {Chen}, Z. and {Chirkin}, D. and {Choi}, S. and {Clark}, B.~A. and {Coleman}, A. and {Collin}, G.~H. and {Connolly}, A. and {Conrad}, J.~M. and {Coppin}, P. and {Correa}, P. and {Cowen}, D.~F. and {Dave}, P. and {De Clercq}, C. and {DeLaunay}, J.~J. and {Delgado}, D. and {Deng}, S. and {Deoskar}, K. and {Desai}, A. and {Desiati}, P. and {de Vries}, K.~D. and {de Wasseige}, G. and {DeYoung}, T. and {Diaz}, A. and {D{\'\i}az-V{\'e}lez}, J.~C. and {Dittmer}, M. and {Domi}, A. and {Dujmovic}, H. and {DuVernois}, M.~A. and {Ehrhardt}, T. and {Eimer}, A. and {Eller}, P. and {Ellinger}, E. and {El Mentawi}, S. and {Els{\"a}sser}, D. and {Engel}, R. and {Erpenbeck}, H. and {Evans}, J. and {Evenson}, P.~A. and {Fan}, K.~L. and {Fang}, K. and {Farrag}, K. and {Fazely}, A.~R. and {Fedynitch}, A. and {Feigl}, N. and {Fiedlschuster}, S. and {Finley}, C. and {Fischer}, L. and {Fox}, D. and {Franckowiak}, A. and {F{\"u}rst}, P. and {Gallagher}, J. and {Ganster}, E. and {Garcia}, A. and {Gerhardt}, L. and {Ghadimi}, A. and {Glaser}, C. and {Glauch}, T. and {Gl{\"u}senkamp}, T. and {Gonzalez}, J.~G. and {Grant}, D. and {Gray}, S.~J. and {Gries}, O. and {Griffin}, S. and {Griswold}, S. and {Groth}, K.~M. and {G{\"u}nther}, C. and {Gutjahr}, P. and {Ha}, C. and {Haack}, C. and {Hallgren}, A. and {Halliday}, R. and {Halve}, L. and {Halzen}, F. and {Hamdaoui}, H. and {Ha Minh}, M. and {Handt}, M. and {Hanson}, K. and {Hardin}, J. and {Harnisch}, A.~A. and {Hatch}, P. and {Haungs}, A. and {H{\"a}u{\ss}ler}, J. and {Helbing}, K. and {Hellrung}, J. and {Hermannsgabner}, J. and {Heuermann}, L. and {Heyer}, N. and {Hickford}, S. and {Hidvegi}, A. and {Hill}, C. and {Hill}, G.~C. and {Hoffman}, K.~D. and {Hori}, S. and {Hoshina}, K. and {Hou}, W. and {Huber}, T. and {Hultqvist}, K. and {H{\"u}nnefeld}, M. and {Hussain}, R. and {Hymon}, K. and {In}, S. and {Ishihara}, A. and {Jacquart}, M. and {Janik}, O. and {Jansson}, M. and {Japaridze}, G.~S. and {Jeong}, M. and {Jin}, M. and {Jones}, B.~J.~P. and {Kamp}, N. and {Kang}, D. and {Kang}, W. and {Kang}, X. and {Kappes}, A. and {Kappesser}, D. and {Kardum}, L. and {Karg}, T. and {Karl}, M. and {Karle}, A. and {Katil}, A. and {Katz}, U. and {Kauer}, M. and {Kelley}, J.~L. and {Khatee Zathul}, A. and {Kheirandish}, A. and {Kiryluk}, J. and {Klein}, S.~R. and {Kochocki}, A. and {Koirala}, R. and {Kolanoski}, H. and {Kontrimas}, T. and {K{\"o}pke}, L. and {Kopper}, C. and {Koskinen}, D.~J. and {Koundal}, P. and {Kovacevich}, M. and {Kowalski}, M. and {Kozynets}, T. and {Krishnamoorthi}, J. and {Kruiswijk}, K. and {Krupczak}, E. and {Kumar}, A. and {Kun}, E. and {Kurahashi}, N. and {Lad}, N. and {Lagunas Gualda}, C. and {Lamoureux}, M. and {Larson}, M.~J. and {Latseva}, S.},
        title = "{Search for 10─1000 GeV Neutrinos from Gamma-Ray Bursts with IceCube}",
      journal = {\apj},
         year = 2024,
        month = apr,
       volume = {964},
       number = {2},
          eid = {126},
        pages = {126},
          doi = {10.3847/1538-4357/ad220b},
archivePrefix = {arXiv},
       eprint = {2312.11515},
 primaryClass = {astro-ph.HE},
       adsurl = {https://ui.adsabs.harvard.edu/abs/2024ApJ...964..126A}
}

@ARTICLE{Pasumarti2,
       author = {{Pasumarti}, Vibhavasu and {Desai}, Shantanu},
        title = "{A stacked search for spatial coincidences between IceCube neutrinos and radio pulsars}",
      journal = {\jcap},
         year = 2024,
        month = apr,
       volume = {2024},
       number = {4},
          eid = {010},
        pages = {010},
          doi = {10.1088/1475-7516/2024/04/010},
archivePrefix = {arXiv},
       eprint = {2306.03427},
 primaryClass = {astro-ph.HE},
       adsurl = {https://ui.adsabs.harvard.edu/abs/2024JCAP...04..010P}
}

@ARTICLE{Hernan,
       author = {{Way}, M.~J. and {Quintana}, H. and {Infante}, L.},
        title = "{The Dynamics of the cD Clusters Abell 119 and Abell 133}",
      journal = {arXiv e-prints},
         year = 1997,
        month = sep,
          eid = {astro-ph/9709036},
        pages = {astro-ph/9709036},
          doi = {10.48550/arXiv.astro-ph/9709036},
archivePrefix = {arXiv},
       eprint = {astro-ph/9709036},
 primaryClass = {astro-ph},
       adsurl = {https://ui.adsabs.harvard.edu/abs/1997astro.ph..9036W}
}

@ARTICLE{Shifa2026,
       author = {{Shifa M.}, Fathima and {Desai}, Shantanu},
        title = "{Search for spatial coincidence between magnetars and IceCube detected neutrinos}",
      journal = {Physics of the Dark Universe},
         year = 2026,
        month = feb,
       volume = {51},
          eid = {102234},
        pages = {102234},
          doi = {10.1016/j.dark.2026.102234},
archivePrefix = {arXiv},
       eprint = {2503.05100},
 primaryClass = {astro-ph.HE},
       adsurl = {https://ui.adsabs.harvard.edu/abs/2026PDU....5102234S}
}

@ARTICLE{Shifa,
       author = {{Shifa M}, Fathima and {Desai}, Shantanu},
        title = "{Search for spatial coincidence between IceCube neutrinos and gamma-ray bright red dwarfs}",
      journal = {Journal of High Energy Astrophysics},
         year = 2025,
        month = jul,
       volume = {47},
          eid = {100366},
        pages = {100366},
          doi = {10.1016/j.jheap.2025.100366},
archivePrefix = {arXiv},
       eprint = {2410.16394},
 primaryClass = {astro-ph.HE},
       adsurl = {https://ui.adsabs.harvard.edu/abs/2025JHEAp..4700366S}
}

@ARTICLE{Chang24,
       author = {{Chang}, Po-Wen and {Zhou}, Bei and {Murase}, Kohta and {Kamionkowski}, Marc},
        title = "{High-energy neutrinos from choked-jet supernovae: Searches and implications}",
      journal = {\prd},
         year = 2024,
        month = may,
       volume = {109},
       number = {10},
          eid = {103041},
        pages = {103041},
          doi = {10.1103/PhysRevD.109.103041},
archivePrefix = {arXiv},
       eprint = {2210.03088},
 primaryClass = {astro-ph.HE},
       adsurl = {https://ui.adsabs.harvard.edu/abs/2024PhRvD.109j3041C}
}

@ARTICLE{IceCubePWN,
       author = {{Aartsen}, M.~G. and {Ackermann}, M. and {Adams}, J. and {Aguilar}, J.~A. and {Ahlers}, M. and {Ahrens}, M. and {Alispach}, C. and {Andeen}, K. and {Anderson}, T. and {Ansseau}, I. and {Anton}, G. and {Arg{\"u}elles}, C. and {Auffenberg}, J. and {Axani}, S. and {Bagherpour}, H. and {Bai}, X. and {Balagopal V.}, A. and {Barbano}, A. and {Barwick}, S.~W. and {Bastian}, B. and {Baum}, V. and {Baur}, S. and {Bay}, R. and {Beatty}, J.~J. and {Becker}, K.-H. and {Becker Tjus}, J. and {BenZvi}, S. and {Berley}, D. and {Bernardini}, E. and {Besson}, D.~Z. and {Binder}, G. and {Bindig}, D. and {Blaufuss}, E. and {Blot}, S. and {Bohm}, C. and {B{\"o}ser}, S. and {Botner}, O. and {B{\"o}ttcher}, J. and {Bourbeau}, E. and {Bourbeau}, J. and {Bradascio}, F. and {Braun}, J. and {Bron}, S. and {Brostean-Kaiser}, J. and {Burgman}, A. and {Buscher}, J. and {Busse}, R.~S. and {Carver}, T. and {Chen}, C. and {Cheung}, E. and {Chirkin}, D. and {Choi}, S. and {Clark}, B.~A. and {Clark}, K. and {Classen}, L. and {Coleman}, A. and {Collin}, G.~H. and {Conrad}, J.~M. and {Coppin}, P. and {Correa}, P. and {Cowen}, D.~F. and {Cross}, R. and {Dave}, P. and {De Clercq}, C. and {DeLaunay}, J.~J. and {Dembinski}, H. and {Deoskar}, K. and {De Ridder}, S. and {Desiati}, P. and {de Vries}, K.~D. and {de Wasseige}, G. and {de With}, M. and {DeYoung}, T. and {Diaz}, A. and {D{\'\i}az-V{\'e}lez}, J.~C. and {Dujmovic}, H. and {Dunkman}, M. and {Dvorak}, E. and {Eberhardt}, B. and {Ehrhardt}, T. and {Eller}, P. and {Engel}, R. and {Evenson}, P.~A. and {Fahey}, S. and {Fazely}, A.~R. and {Felde}, J. and {Filimonov}, K. and {Finley}, C. and {Fox}, D. and {Franckowiak}, A. and {Friedman}, E. and {Fritz}, A. and {Gaisser}, T.~K. and {Gallagher}, J. and {Ganster}, E. and {Garrappa}, S. and {Gerhardt}, L. and {Ghorbani}, K. and {Glauch}, T. and {Gl{\"u}senkamp}, T. and {Goldschmidt}, A. and {Gonzalez}, J.~G. and {Grant}, D. and {Gr{\'e}goire}, T. and {Griffith}, Z. and {Griswold}, S. and {G{\"u}nder}, M. and {G{\"u}nd{\"u}z}, M. and {Haack}, C. and {Hallgren}, A. and {Halliday}, R. and {Halve}, L. and {Halzen}, F. and {Hanson}, K. and {Haungs}, A. and {Hebecker}, D. and {Heereman}, D. and {Heix}, P. and {Helbing}, K. and {Hellauer}, R. and {Henningsen}, F. and {Hickford}, S. and {Hignight}, J. and {Hill}, G.~C. and {Hoffman}, K.~D. and {Hoffmann}, R. and {Hoinka}, T. and {Hokanson-Fasig}, B. and {Hoshina}, K. and {Huang}, F. and {Huber}, M. and {Huber}, T. and {Hultqvist}, K. and {H{\"u}nnefeld}, M. and {Hussain}, R. and {In}, S. and {Iovine}, N. and {Ishihara}, A. and {Jansson}, M. and {Japaridze}, G.~S. and {Jeong}, M. and {Jero}, K. and {Jones}, B.~J.~P. and {Jonske}, F. and {Joppe}, R. and {Kang}, D. and {Kang}, W. and {Kappes}, A. and {Kappesser}, D. and {Karg}, T. and {Karl}, M. and {Karle}, A. and {Katz}, U. and {Kauer}, M. and {Kellermann}, M. and {Kelley}, J.~L. and {Kheirandish}, A. and {Kim}, J. and {Kintscher}, T. and {Kiryluk}, J. and {Kittler}, T. and {Klein}, S.~R. and {Koirala}, R. and {Kolanoski}, H. and {K{\"o}pke}, L. and {Kopper}, C. and {Kopper}, S. and {Koskinen}, D.~J. and {Kowalski}, M. and {Krings}, K. and {Kr{\"u}ckl}, G. and {Kulacz}, N. and {Kurahashi}, N. and {Kyriacou}, A. and {Lanfranchi}, J.~L. and {Larson}, M.~J. and {Lauber}, F. and {Lazar}, J.~P. and {Leonard}, K. and {Leszczy{\'n}ska}, A. and {Liu}, Q.~R. and {Lohfink}, E. and {Lozano Mariscal}, C.~J. and {Lu}, L. and {Lucarelli}, F. and {Ludwig}, A. and {L{\"u}nemann}, J. and {Luszczak}, W. and {Lyu}, Y. and {Ma}, W.~Y. and {Madsen}, J. and {Maggi}, G. and {Mahn}, K.~B.~M. and {Makino}, Y. and {Mallik}, P. and {Mallot}, K. and {Mancina}, S. and {Mari{\textcommabelow s}}, I.~C. and {Maruyama}, R. and {Mase}, K.},
        title = "{IceCube Search for High-energy Neutrino Emission from TeV Pulsar Wind Nebulae}",
      journal = {\apj},
         year = 2020,
        month = aug,
       volume = {898},
       number = {2},
          eid = {117},
        pages = {117},
          doi = {10.3847/1538-4357/ab9fa0},
archivePrefix = {arXiv},
       eprint = {2003.12071},
 primaryClass = {astro-ph.HE},
       adsurl = {https://ui.adsabs.harvard.edu/abs/2020ApJ...898..117A}
}

@ARTICLE{DesaiFRB,
       author = {{Desai}, Shantanu},
        title = "{A test of spatial coincidence between CHIME FRBs and IceCube TeV energy neutrinos}",
      journal = {Journal of Physics G Nuclear Physics},
         year = 2023,
        month = jan,
       volume = {50},
       number = {1},
          eid = {015201},
        pages = {015201},
          doi = {10.1088/1361-6471/aca03b},
archivePrefix = {arXiv},
       eprint = {2112.13820},
 primaryClass = {astro-ph.HE},
       adsurl = {https://ui.adsabs.harvard.edu/abs/2023JPhG...50a5201D}
}

@ARTICLE{Beacom,
       author = {{Zhou}, Bei and {Beacom}, John F.},
        title = "{Dimuons in neutrino telescopes: New predictions and first search in IceCube}",
      journal = {\prd},
         year = 2022,
        month = may,
       volume = {105},
       number = {9},
          eid = {093005},
        pages = {093005},
          doi = {10.1103/PhysRevD.105.093005},
archivePrefix = {arXiv},
       eprint = {2110.02974},
 primaryClass = {hep-ph},
       adsurl = {https://ui.adsabs.harvard.edu/abs/2022PhRvD.105i3005Z}
}

@ARTICLE{Montaruli,
       author = {{Braun}, Jim and {Dumm}, Jon and {De Palma}, Francesco and {Finley}, Chad and {Karle}, Albrecht and {Montaruli}, Teresa},
        title = "{Methods for point source analysis in high energy neutrino telescopes}",
      journal = {Astroparticle Physics},
         year = 2008,
        month = may,
       volume = {29},
       number = {4},
        pages = {299-305},
          doi = {10.1016/j.astropartphys.2008.02.007},
archivePrefix = {arXiv},
       eprint = {0801.1604},
 primaryClass = {astro-ph},
       adsurl = {https://ui.adsabs.harvard.edu/abs/2008APh....29..299B}
}

@ARTICLE{Mattox96,
       author = {{Mattox}, J.~R. and {Bertsch}, D.~L. and {Chiang}, J. and {Dingus}, B.~L. and {Digel}, S.~W. and {Esposito}, J.~A. and {Fierro}, J.~M. and {Hartman}, R.~C. and {Hunter}, S.~D. and {Kanbach}, G. and {Kniffen}, D.~A. and {Lin}, Y.~C. and {Macomb}, D.~J. and {Mayer-Hasselwander}, H.~A. and {Michelson}, P.~F. and {von Montigny}, C. and {Mukherjee}, R. and {Nolan}, P.~L. and {Ramanamurthy}, P.~V. and {Schneid}, E. and {Sreekumar}, P. and {Thompson}, D.~J. and {Willis}, T.~D.},
        title = "{The Likelihood Analysis of EGRET Data}",
      journal = {\apj},
         year = 1996,
        month = apr,
       volume = {461},
        pages = {396},
          doi = {10.1086/177068},
       adsurl = {https://ui.adsabs.harvard.edu/abs/1996ApJ...461..396M}
}

@ARTICLE{Murase,
       author = {{Murase}, Kohta and {Beacom}, John F.},
        title = "{Galaxy clusters as reservoirs of heavy dark matter and high-energy cosmic rays: constraints from neutrino observations}",
      journal = {\jcap},
         year = 2013,
        month = feb,
       volume = {2013},
       number = {2},
          eid = {028},
        pages = {028},
          doi = {10.1088/1475-7516/2013/02/028},
archivePrefix = {arXiv},
       eprint = {1209.0225},
 primaryClass = {astro-ph.HE},
       adsurl = {https://ui.adsabs.harvard.edu/abs/2013JCAP...02..028M}
}

@ARTICLE{Nuza,
       author = {{Supanitsky}, A.~D. and {Nuza}, S.~E.},
        title = "{Exploring the role of accretion shocks in galaxy clusters as sources of ultrahigh-energy cosmic rays}",
      journal = {\prd},
         year = 2026,
        month = feb,
       volume = {113},
       number = {4},
          eid = {043008},
        pages = {043008},
          doi = {10.1103/xz32-7pk7},
archivePrefix = {arXiv},
       eprint = {2601.18411},
 primaryClass = {astro-ph.HE},
       adsurl = {https://ui.adsabs.harvard.edu/abs/2026PhRvD.113d3008S}
}

@ARTICLE{Allen2011,
       author = {{Allen}, Steven W. and {Evrard}, August E. and {Mantz}, Adam B.},
        title = "{Cosmological Parameters from Observations of Galaxy Clusters}",
      journal = {\araa},
         year = 2011,
        month = sep,
       volume = {49},
       number = {1},
        pages = {409-470},
          doi = {10.1146/annurev-astro-081710-102514},
archivePrefix = {arXiv},
       eprint = {1103.4829},
 primaryClass = {astro-ph.CO},
       adsurl = {https://ui.adsabs.harvard.edu/abs/2011ARA&A..49..409A}
}

@ARTICLE{Miyatake25,
       author = {{Miyatake}, Hironao},
        title = "{Cosmology with Galaxy Clusters}",
      journal = {arXiv e-prints},
         year = 2025,
        month = may,
          eid = {arXiv:2505.07697},
        pages = {arXiv:2505.07697},
          doi = {10.48550/arXiv.2505.07697},
archivePrefix = {arXiv},
       eprint = {2505.07697},
 primaryClass = {astro-ph.CO},
       adsurl = {https://ui.adsabs.harvard.edu/abs/2025arXiv250507697M}
}

@ARTICLE{Bohringer16,
       author = {{B{\"o}hringer}, Hans and {Chon}, Gayoung},
        title = "{Constraints on neutrino masses from the study of the nearby large-scale structure and galaxy cluster counts}",
      journal = {Modern Physics Letters A},
         year = 2016,
        month = jul,
       volume = {31},
       number = {21},
          eid = {1640008},
        pages = {1640008},
          doi = {10.1142/S0217732316400083},
archivePrefix = {arXiv},
       eprint = {1610.02855},
 primaryClass = {astro-ph.CO},
       adsurl = {https://ui.adsabs.harvard.edu/abs/2016MPLA...3140008B}
}

@ARTICLE{Hennig,
       author = {{Hennig}, C. and {Mohr}, J.~J. and {Zenteno}, A. and {Desai}, S. and {Dietrich}, J.~P. and {Bocquet}, S. and {Strazzullo}, V. and {Saro}, A. and {Abbott}, T.~M.~C. and {Abdalla}, F.~B. and {Bayliss}, M. and {Benoit-L{\'e}vy}, A. and {Bernstein}, R.~A. and {Bertin}, E. and {Brooks}, D. and {Capasso}, R. and {Capozzi}, D. and {Carnero}, A. and {Carrasco Kind}, M. and {Carretero}, J. and {Chiu}, I. and {D'Andrea}, C.~B. and {daCosta}, L.~N. and {Diehl}, H.~T. and {Doel}, P. and {Eifler}, T.~F. and {Evrard}, A.~E. and {Fausti-Neto}, A. and {Fosalba}, P. and {Frieman}, J. and {Gangkofner}, C. and {Gonzalez}, A. and {Gruen}, D. and {Gruendl}, R.~A. and {Gupta}, N. and {Gutierrez}, G. and {Honscheid}, K. and {Hlavacek-Larrondo}, J. and {James}, D.~J. and {Kuehn}, K. and {Kuropatkin}, N. and {Lahav}, O. and {March}, M. and {Marshall}, J.~L. and {Martini}, P. and {McDonald}, M. and {Melchior}, P. and {Miller}, C.~J. and {Miquel}, R. and {Neilsen}, E. and {Nord}, B. and {Ogando}, R. and {Plazas}, A.~A. and {Reichardt}, C. and {Romer}, A.~K. and {Rozo}, E. and {Rykoff}, E.~S. and {Sanchez}, E. and {Santiago}, B. and {Schubnell}, M. and {Sevilla-Noarbe}, I. and {Smith}, R.~C. and {Soares-Santos}, M. and {Sobreira}, F. and {Stalder}, B. and {Stanford}, S.~A. and {Suchyta}, E. and {Swanson}, M.~E.~C. and {Tarle}, G. and {Thomas}, D. and {Vikram}, V. and {Walker}, A.~R. and {Zhang}, Y.},
        title = "{Galaxy populations in massive galaxy clusters to z = 1.1: colour distribution, concentration, halo occupation number and red sequence fraction}",
      journal = {\mnras},
         year = 2017,
        month = jun,
       volume = {467},
       number = {4},
        pages = {4015-4035},
          doi = {10.1093/mnras/stx175},
archivePrefix = {arXiv},
       eprint = {1604.00988},
 primaryClass = {astro-ph.GA},
       adsurl = {https://ui.adsabs.harvard.edu/abs/2017MNRAS.467.4015H}
}

@ARTICLE{Brunetti2014,
       author = {{Brunetti}, Gianfranco and {Jones}, Thomas W.},
        title = "{Cosmic Rays in Galaxy Clusters and Their Nonthermal Emission}",
      journal = {International Journal of Modern Physics D},
         year = 2014,
        month = mar,
       volume = {23},
       number = {4},
          eid = {1430007-98},
        pages = {1430007-98},
          doi = {10.1142/S0218271814300079},
archivePrefix = {arXiv},
       eprint = {1401.7519},
 primaryClass = {astro-ph.CO},
       adsurl = {https://ui.adsabs.harvard.edu/abs/2014IJMPD..2330007B}
}

@ARTICLE{TA,
       author = {{Telescope Array Collaboration} and {Abbasi}, R.~U. and {Abu-Zayyad}, T. and {Allen}, M. and {Arai}, Y. and {Arimura}, R. and {Barcikowski}, E. and {Belz}, J.~W. and {Bergman}, D.~R. and {Blake}, S.~A. and {Buckland}, I. and {Cady}, R. and {Cheon}, B.~G. and {Chiba}, J. and {Chikawa}, M. and {Fujii}, T. and {Fujisue}, K. and {Fujita}, K. and {Fujiwara}, R. and {Fukushima}, M. and {Fukushima}, R. and {Furlich}, G. and {Globus}, N. and {Gonzalez}, R. and {Hanlon}, W. and {Hayashi}, M. and {Hayashida}, N. and {Hibino}, K. and {Higuchi}, R. and {Honda}, K. and {Ikeda}, D. and {Inadomi}, T. and {Inoue}, N. and {Ishii}, T. and {Ito}, H. and {Ivanov}, D. and {Iwakura}, H. and {Iwasaki}, A. and {Jeong}, H.~M. and {Jeong}, S. and {Jui}, C.~C.~H. and {Kadota}, K. and {Kakimoto}, F. and {Kalashev}, O. and {Kasahara}, K. and {Kasami}, S. and {Kawai}, H. and {Kawakami}, S. and {Kawana}, S. and {Kawata}, K. and {Kharuk}, I. and {Kido}, E. and {Kim}, H.~B. and {Kim}, J.~H. and {Kim}, J.~H. and {Kim}, M.~H. and {Kim}, S.~W. and {Kimura}, Y. and {Kishigami}, S. and {Kubota}, Y. and {Kurisu}, S. and {Kuzmin}, V. and {Kuznetsov}, M. and {Kwon}, Y.~J. and {Lee}, K.~H. and {Lubsandorzhiev}, B. and {Lundquist}, J.~P. and {Machida}, K. and {Matsumiya}, H. and {Matsuyama}, T. and {Matthews}, J.~N. and {Mayta}, R. and {Minamino}, M. and {Mukai}, K. and {Myers}, I. and {Nagataki}, S. and {Nakai}, K. and {Nakamura}, R. and {Nakamura}, T. and {Nakamura}, T. and {Nakamura}, Y. and {Nakazawa}, A. and {Nishio}, E. and {Nonaka}, T. and {Oda}, H. and {Ogio}, S. and {Ohnishi}, M. and {Ohoka}, H. and {Oku}, Y. and {Okuda}, T. and {Omura}, Y. and {Ono}, M. and {Onogi}, R. and {Oshima}, A. and {Ozawa}, S. and {Park}, I.~H. and {Potts}, M. and {Pshirkov}, M.~S. and {Remington}, J. and {Rodriguez}, D.~C. and {Rubtsov}, G.~I. and {Ryu}, D. and {Sagawa}, H. and {Sahara}, R. and {Saito}, Y. and {Sakaki}, N. and {Sako}, T. and {Sakurai}, N. and {Sano}, K. and {Sato}, K. and {Seki}, T. and {Sekino}, K. and {Shah}, P.~D. and {Shibasaki}, Y. and {Shibata}, F. and {Shibata}, N. and {Shibata}, T. and {Shimodaira}, H. and {Shin}, B.~K. and {Shin}, H.~S. and {Shinto}, D. and {Smith}, J.~D. and {Sokolsky}, P. and {Sone}, N. and {Stokes}, B.~T. and {Stroman}, T.~A. and {Takagi}, Y. and {Takahashi}, Y. and {Takamura}, M. and {Takeda}, M. and {Takeishi}, R. and {Taketa}, A. and {Takita}, M. and {Tameda}, Y. and {Tanaka}, H. and {Tanaka}, K. and {Tanaka}, M. and {Tanoue}, Y. and {Thomas}, S.~B. and {Thomson}, G.~B. and {Tinyakov}, P. and {Tkachev}, I. and {Tokuno}, H. and {Tomida}, T. and {Troitsky}, S. and {Tsuda}, R. and {Tsunesada}, Y. and {Uchihori}, Y. and {Udo}, S. and {Uehama}, T. and {Urban}, F. and {Wong}, T. and {Yamamoto}, M. and {Yamazaki}, K. and {Yang}, J. and {Yashiro}, K. and {Yoshida}, F. and {Yoshioka}, Y. and {Zhezher}, Y. and {Zundel}, Z.},
        title = "{Indications of a Cosmic Ray Source in the Perseus-Pisces Supercluster}",
      journal = {arXiv e-prints},
         year = 2021,
        month = oct,
          eid = {arXiv:2110.14827},
        pages = {arXiv:2110.14827},
          doi = {10.48550/arXiv.2110.14827},
archivePrefix = {arXiv},
       eprint = {2110.14827},
 primaryClass = {astro-ph.HE},
       adsurl = {https://ui.adsabs.harvard.edu/abs/2021arXiv211014827T}
}

@ARTICLE{Fermi2010,
       author = {{Ackermann}, M. and {Ajello}, M. and {Allafort}, A. and {Baldini}, L. and {Ballet}, J. and {Barbiellini}, G. and {Bastieri}, D. and {Bechtol}, K. and {Bellazzini}, R. and {Blandford}, R.~D. and others},
        title = "{GeV Gamma-ray Flux Upper Limits from Clusters of Galaxies}",
      journal = {\apjl},
         year = 2010,
        month = jul,
       volume = {717},
       number = {1},
        pages = {L71-L78},
          doi = {10.1088/2041-8205/717/1/L71},
archivePrefix = {arXiv},
       eprint = {1006.0748},
 primaryClass = {astro-ph.HE},
       adsurl = {https://ui.adsabs.harvard.edu/abs/2010ApJ...717L..71A}
}
\appendix
\section{Extended Source Analysis for Abell 119}
\rthis{To complement the point source analysis, we carry out an extended source analysis, following the prescription in  ~\cite{Icecubeextended}. The main difference from the point source analysis is that $\sigma_i$ in Eq.~\ref{eq:L} is replaced by $\sqrt{\sigma_i^2+\sigma_s^2}$, where $\sigma_s$ is the angular extent of the source. 
We approximate the spatial morphology of Abell 119 by a Gaussian source with width $\sigma_s=\theta_{200}$, where  $\theta_{200}=0.6^{\circ}$. The rest of the analysis is the same as the point source analysis carried out in the main manuscript. We find the maximum value of $\hat{n}_s$ equal to 28.9, with maximum TS value of 2.3. Therefore although, the maximum TS value is slightly larger than the point analysis, it is still not statistically significant. The 90\% c.l. upper limit on $\hat{n}_s$ is equal to 55.1. The corresponding differential muon  neutrino energy flux limit at a pivot energy of 100 TeV for a spectral index of 2.0 is given by 
$E^2 \phi_{\nu} =  2.76 \times 10^{-9} ~\mathrm{GeV}~\mathrm{cm}^{-2}~\mathrm{s}^{-1}$. Therefore, the limit is degraded by a factor of  two compared to the point source analysis.}
\end{document}